\documentstyle[12pt]{article}

\begin{document}

\begin{center}
  \large {\bf Projection Postulate and Atomic Quantum Zeno Effect}

\vspace*{.5cm}

   Almut Beige and Gerhard C. Hegerfeldt\\ [.5cm]
\normalsize
   Institute for Theoretical Physics\cite{email}\\
   University of G\"ottingen\\
   Bunsenstr. 9\\
   D 37073 G\"ottingen, Germany
\end{center}

\vspace*{1cm}

\begin{abstract}
The projection postulate has been used to predict a slow-down of the
time evolution of the state of a system
under rapidly repeated measurements, and ultimately a
freezing of the state. To test this so-called quantum Zeno effect an
experiment  was performed by Itano et al.
(Phys. Rev. A {\bf 41}, 2295 (1990)) in which
an atomic-level measurement was realized by means of
a short laser pulse.
The relevance of the results
has given rise to controversies in the literature. In
particular the projection postulate and its applicability
in this experiment have been cast into doubt.
In this paper we show analytically that for a wide range of
parameters such a short laser pulse
acts as an {\em effective} level measurement to which the usual projection
postulate applies with high accuracy. The corrections to the ideal
reductions and their accumulation over $n$  pulses are calculated.
Our conclusion is that the projection postulate
is an excellent pragmatic tool for a quick and simple understanding of the
slow-down of time evolution in experiments of this type.
However,  corrections have to be included, and an actual freezing
does not seem possible because of the finite duration of
measurements.\\[.25cm]
PACS numbers 03.65.Bz; 42.50.-p; 32.80.-t
\end{abstract}
\vspace*{0.5cm}

\noindent {\bf 1. Introduction}

\vspace*{1cm}

The so-called quantum Zeno effect (QZE) \cite{MiSu} is a theoretical
prediction for the behavior of a system under rapidly repeated measurements
at times $\Delta t$ apart. It is based on usual quantum theory and on the
concept of instantaneous measurements together with ensuing state reductions
according to the projection postulate of von Neumann and L\"uders \cite{N}.
The predictions are:
\begin{enumerate}
\item Impediment and slow-down of the time development of the system due to
repeated measurements.
\item Freezing of the state for $\Delta t \rightarrow 0$, i.e. in the
limit of continuous measurements.
\end{enumerate}
The underlying reason for this can be traced to the fact that for short
enough times transition probabilities grow quadratically with time.
If, in a given time interval $T$, one performs $n$ measurements at
times $\Delta t = T/n$ apart, then the probability to find an
orthonormal  state is at most
proportional to $n (\Delta t)^2 = T^2/n$, which goes to zero for $n
\rightarrow \infty$ or $\Delta t \rightarrow 0$.

Properties 1 and 2 may be taken as a definition of the QZE, as for instance
in Refs. \cite{Wine,B15,B16,B4}. A slightly different definition is used
in Ref. \cite{B2} where essentially only property 2 is used.

An experiment to test the QZE for atomic systems has been performed by
Itano et al. \cite{Wine}, following a suggestion by Cook \cite{B2}.
In the experiment a large number $N$ of ions was stored in a Penning trap
(see Fig. 1 for the relevant level structure, a V configuration;
level 2 is (meta-)stable). The time development is given by a so-called
$\pi$ pulse \cite{pulse} of length $T_{\pi}$, tuned to the 1
- 2 transition frequency. A very short pulse of a probe laser couples level
1 with an auxiliary third level, and this is regarded as a
measurement to which the projection postulate is applied as follows.
It is assumed that an atom is either in level 1 or level 2, depending
on whether or not it has emitted photons during the probe pulse
\cite{obvious}. At the end of the $\pi$ pulse
the final level populations for up to
64 probe pulses ("measurements") during a $\pi$ pulse (see Fig. 2)
are determined
and found to be in agreement with
the quantum Zeno predictions \cite{compl}.

The interpretation of this experiment, as to whether it does or does not
bear on the QZE, has been controversial in the literature. Some have
hailed the results as a dramatic verification of the QZE, others
argue that they are unrelated to it
\cite{B15,B16,Peres,crit1,crit2,crit5,crit4,Gag,bloch}. In particular the
use of the
projection postulate and reduction of the wave-function have been criticized,
and the very existence of the QZE has been cast into doubt. It is no
exaggeration  to say that the QZE has aroused tremendous interest in the
literature \cite{interest,Power,B36}.

It was pointed out in Refs. \cite{B15,B16,crit5,crit4,Gag,bloch} that these
results could be understood without any recourse to the projection postulate
or to the QZE. One can simply incorporate the short probe pulses in the
dynamics by an appropriate term in the Hamiltonian or in the optical
Bloch equations. A numerical solution of these Bloch equations should
then yield the experimental result, and indeed they do to good
agreement. The projection postulate does not seem to be needed. Is it
also {\it incorrect} to use it here?

This is one of the main two questions we are  going to address in this
paper. First we give a
justification of the projection postulate as a useful and approximate
{\em technical} tool in experimental situations of the general setup
considered in Ref. \cite{Wine} for a wide range of parameters.
Our motivation differs somewhat from
that of Refs. \cite{B15,B16,crit5,crit4,Gag,bloch}, although there
is no contradiction, in particular not to Refs. \cite{B16,Gag}. In
Ref. \cite{B16} the ensemble density matrix is calculated by
adiabatic Bloch equation techniques for
a single probe pulse and  found to be nearly diagonal.
But this contains no information
about the outcome for individual atoms and subensembles and does not
prove that atoms with or without photon emissions are in $|1
\rangle$ and $|2 \rangle$, respectively. The same technique is used
in Ref. \cite{Gag} to study the parameter domain of "good"
measurements with a resulting diagonal density matrix \cite{Gag2}.
We, however, are concerned  first with selective measurements, namely
with the states of atoms at the end of a probe pulse with
or without photon emissions.
These states are explicitly determined for the first time and found to be
close, but not identical, to $|1
\rangle$ and $|2 \rangle$ \cite{corr}.
Thus a probe pulse can indeed be regarded
as accomplishing a highly accurate -- but not  perfect -- reduction to and
measurement of  levels 1 and 2  in experiments  of the type of
Ref. \cite{Wine} for a wide range of parameters. As a second question
we discuss the cumulative effect of the deviations from  ideal measurements
-- i.e. measurements which can be described   by the projection
postulate -- over $n$ probe pulses for the density matrix
and  determine the resulting level
populations at the end of the $\pi$ pulse. We exhibit
an explicit $n$ dependence
of the corrections to the treatment by the projection postulate. This
is also a new result.

For a probe pulse to be an effective measurement some obvious requirements
have to be fulfilled. First of all, its duration, $\tau_{\rm p}$,
should be very short compared to the duration $T_{\rm \pi}$ of the $\pi $
pulse,
\begin{equation}\label{1.10}
\tau_{\rm p} \ll T_{\rm_\pi}~.
\end{equation}
Furthermore, the probe pulse cannot be too short or too weak, because
it should produce fluorescence photons from level 1 with high
certainty.
In addition, at the end of a probe pulse any population in the auxiliary
level 3 should decay to level 1 extremely rapidly. This means
that $A_3^{-1}$, the inverse Einstein coefficient of level 3, must be
tiny compared to the time between two probe pulses,
\begin{equation}\label{1.11}
A_3^{-1} \ll \frac{T_{\rm \pi}}{n} - \tau_{\rm p}~ .
\end{equation}
The first two conditions lead to a restriction on two parameters,
namely \cite{proof}
\begin{equation}\label{1.15}
\epsilon_{\rm p} \equiv ~\frac{A_3 \Omega_2}{\Omega_3^2}~\ll 1~~~~
\mbox{and}~~~~\epsilon_{\rm R} \equiv ~\frac{\Omega_2}{\Omega_3}~\ll
1~.
\end{equation}
In Sections 4 and 5 we will use the conditions
\begin{equation}\label{bed}
\epsilon_{\rm p} \equiv ~\frac{A_3 \Omega_2}{\Omega_3^2}~\ll 1~~~~
\mbox{and}~~~~\epsilon_{\rm d} \equiv ~\frac{\Omega_3^2}{A_3^2}~\ll
1~.
\end{equation}
Then the condition $\epsilon_{\rm R} \ll 1$ is automatically
fulfilled since $\epsilon_{\rm R} = \epsilon_{\rm p} \epsilon_{\rm
d}^{1/2}$. This parameter regime is compatible with the one
investigated in Ref. \cite{Gag}.
In the experiment of Itano et al. \cite{Wine} $\epsilon_{\rm R}$ is about
$6.5\cdot10^{-6}$, $\epsilon_{\rm p}$ about $4.1\cdot10^{-4}$, and
$\epsilon_{\rm d}$ about $2.5\cdot10^{-4}$ \cite{general}.

In our paper we mainly use   the  quantum jump approach
(quantum trajectories) \cite{HeWi,He,QT}, and
because of its   inherent simplicity
the analysis can be carried out analytically here
for  a wide range of experimental
parameters. This approach is equivalent to the
the Monte-Carlo wave function approach \cite{MC}.
The ensemble of all trajectories satisfies the usual Bloch equations and
is often used to obtain numerical solutions of the latter through
numerical simulations. The quantum jump approach
deals with pure  states for single atoms instead of density matrices
for ensembles \cite{ensemble}, which in the
present case   means a simplification from nine components to three,
and it describes the time development of single
atoms between photon emissions by a simple
reduced Hamiltonian and by a "jump" to a reset state (in this
paper the ground state) at a photon detection \cite{jump}.
It allows an intuitive understanding of
the processes involved.

The simple
derivation of the quantum jump approach  outlined  in Section 2 uses
the projection postulate and reductions
as a technical tool for photon detections (not for atomic measurements!),
and it might seem that one uses this postulate for an investigation of
itself. However, as explained elsewhere \cite{HeWi} the
use of reductions in the derivation of  quantum trajectories
is not necessary and can be avoided. Moreover,
the photon detections happen on a much shorter time scale than all
times considered in the experiment, and for such  detections
the projection postulate has been a reliable tool in the past.

In Section 3 we consider the simple
case of a probe pulse with the $\pi$ pulse
temporarily switched off. In this example it is very easy to see how
the probe pulse acts on a state $\alpha_1 | 1 \rangle+\alpha_2 |2 \rangle$
of a single atom and that it effectively leaves it either in $| 1 \rangle$
or in $| 2\rangle $, with probability $| \alpha_1 |^2$
and $|\alpha_2 |^2$, respectively,
depending on whether the atom has emitted photons or not. The
small deviations from an ideal measurement with perfect state
reduction are determined.

In Section 4 we consider the case of a single probe pulse with the
$\pi$ pulse switched on. Complications arise now since the $\pi$
pulse causes a small additional transition between 1 and 2.
In Section 5 the case of $n$ probe pulses during a single $\pi$
pulse and
the build-up of the corrections are  considered.
It turns out that one gets an excellent
approximation for the level populations
if in the results for the idealized case one takes the
finite duration of the probe pulse into account by neglecting the action
of the $\pi$ pulse during this time (see Fig. 3). This simply means that
$T_{\rm \pi} /n$ is replaced by
$\Delta T~ =~T_{\rm \pi} / n - \tau_{\rm p}$.
At the end of Section 5 we
compare our analytical results with
those of a numerical solution of the optical Bloch equations
for the parameters of the experiment. The agreement is amazing.

In the last section we discuss our result and their significance for
the role of the projection postulate in the so-called quantum Zeno effect.
Our main conclusion is that, although the projection postulate is
not necessary here, it is a useful  tool for a fairly accurate
description of the measurements involved. It is  useful since it
gives a quick intuitive understanding of the physical situation
-- without having to solve unwieldy Bloch equation.
Insofar the projection postulate should not be dismissed
out of hand. However, it is only approximately valid, because in
practice realistic measurements are never ideal nor instantaneous.
Therefore the first part of the Zeno effect -- impediment and slow-down of
the time development -- can legitimately be understood by the projection
postulate, but the second part -- the freezing of the state  -- is in our
opinion an over-idealization.

\vspace*{1cm}

\noindent{\bf 2. The quantum jump approach in quantum optics. Quantum
trajectories}
\\[1.cm]
In this section we briefly summarize the quantum jump approach used
in the subsequent sections. The reader familiar with it can proceed
directly to Eqs. (\ref{2.12}) - (\ref{2.17}). The idea is to describe the
radiating atom between photon detections by a reduced (or effective)
time evolution operator giving the time development under the
condition that no photon has been detected \cite{HeWi}. After a photon
detection one has to reset the atom to the reset state ("jump"),
with ensuing
reduced time development, and so on.
For a driven  system with many emissions one then obtains a
stochastic path, called a quantum trajectory \cite{QT}.
The general reset
state has been determined in Ref. \cite{He}. For a V system as considered
in this paper the reset state after an emission is  the ground state.
The reduced
time development together with the reset state provide a complete
stochastic description
of the time development of the atom \cite{He}. Starting with this description
one can then derive the Bloch equations describing an ensemble of radiating
atoms \cite{HeWi,He}. In fact both approaches, quantum jumps and Bloch
equations, are possible and equivalent ways to describe the time evolution
of an ensemble of  fluorescing atoms, but the former is also easy to
apply to the emission behavior of a single atom.

We now indicate how to determine the reduced time development
operator for the V system. To be sure that
no photon has been detected in a time interval one may imagine
 measurements on the radiation field in a rapid succession
at times $\Delta t$ apart, at $t_1 < \cdots < t_m = t$ say.
If in all these measurements no photons are found the state of the
atom at time $t$ is, by the von Neumann - L\"uders projection postulate
\cite{red},
\begin{eqnarray}
|\phi(t)\rangle &=& |0_{ph}\rangle\langle 0_{ph}|
U(t_m,t_{m-1}) |0_{ph}\rangle\langle 0_{ph}|
\ldots\nonumber\\
& &\hspace*{2cm} \ldots|0_{ph}\rangle\langle
0_{ph}|U(t_1,t_0)|0_{ph}\rangle|\psi\rangle \label{2.1}\\
&=& |0_{ph}\rangle U_{\rm red}(t,t_0)|\psi\rangle .\label{2.1a}
\end{eqnarray}
where $|0_{ph}\rangle$ is the vacuum state,
$|0_{ph}\rangle |\psi \rangle$ the initial state, $U$ the complete time
development operator, and
where the second equality serves as a definition of the reduced time
evolution operator $U_{\rm red}(t,t_0)$, which acts on atomic states.
The time difference between successive measurements has to be chosen
short enough to be able to say that at most one photon has been
detected
in this interval. On the other hand this time difference has to be
longer than the inverse transition
frequencies \cite{HeWi}.
Under these
assumptions one can calculate Eq. (\ref{2.1}) by means of perturbation
theory.
With $U_{\rm red}(t,t_0)$ one has a simple
expression for the probability $P_0(t)$ that no photon is
detected in the interval $[0,t]$ if the atom is in the state
$|\psi\rangle$ at $t=0$,
\begin{eqnarray}
P_0(t) &=& ||~|\phi(t)\rangle||^2\nonumber\\
       &=& ||U_{\rm red}(t,t_0)|\psi\rangle||^2 ~.
\label{2.2a}
\end{eqnarray}

We now apply this to the V system depicted in Fig. 1. In this system
the upper levels 2 and 3 couple to a
common ground level 1, with Einstein coefficients $A_2 $ and $A_3$.
Later we will consider level 2 to be stable and put $A_2 = 0$.
We assume here that
$\omega_{32}~\equiv~\omega_{3} - \omega_{2} $ is in the optical
range, i.e. not too small. For simplicity we consider zero detunings
of the driving fields, whose (real)
Rabi frequencies are denoted by $\Omega_2$
and $\Omega_3$, respectively. The Hamiltonian in
rotating wave approximation is given by \cite{Loudon}
\begin{eqnarray}
H = H^0_A + H^0_F +\sum_{i=2}^3 \hbar \{{1\over 2} \Omega_i |i\rangle\langle
1|e^{-i\omega_it} + {\rm h.c.}\} +
\sum_{i=2}^3\sum_{{\bf k}\lambda}\hbar\left\{g_{i{\bf k}\lambda}
a_{{\bf k}\lambda}|i\rangle\langle 1| + {\rm h.c.} \right\}\nonumber
\end{eqnarray}
where
\begin{eqnarray}
g_{i{\bf k}\lambda}=i\,e {\bf D}_{i1}\cdot {\bf \epsilon}_{{\bf
k}\lambda} \left(\omega_{{\bf k}\lambda}/\left(
2\epsilon_0\hbar V\right) \right)^{1/2}~,
\end{eqnarray}
with the transition dipole moment ${\bf D}_{i1}=\langle i|{\bf X}|
1\rangle$ and ${\bf \epsilon}_{{\bf k}\lambda}$ the polarization
vector. $V$ is the quan\-ti\-za\-tion volume, later taken to in\-finity. Going
over to an interaction picture with respect to
\begin{equation}
H_0 = H^0_A + H^0_F \label{2.5}
\end{equation}
one has
\begin{equation}
H_I(t) = \sum_{i=2}^3 \hbar \{{1\over 2} \Omega_i |i\rangle\langle
1| + {\rm h.c.}\} + \sum_{i=2}^3\sum_{{\bf k}\lambda}\hbar\left\{g_{i{\bf
k}\lambda}
a_{{\bf k}\lambda}|i\rangle\langle 1|
e^{i(\omega_{i}-\omega_{{\bf k}\lambda})t} + {\rm h.c.} \right\}\,
.\label{2.6}
\end{equation}
By $U_I$ we denote the corresponding time development operator.
With $\Delta t = t_i-t_{i-1}$ in the range given above we can calculate in
second order perturbation theory the time evolution under the
condition
that no photon has been detected. From
\begin{eqnarray}
& &\langle 0_{ph}|U_I(t_i,t_{i-1})|0_{ph}\rangle
\nonumber\\
&=&1_A -\frac{i}{\hbar}\int_{t_{i-1}}^{t_i} dt' \langle
0_{ph}|H_I(t')|0_{ph}\rangle
- \frac{1}{\hbar^2}\int_{t_{i-1}}^{t_i} dt'\int_{t_{i-
1}}^{t'}\!\! dt''
\langle 0_{ph}|H_I(t')H_I(t'')|0_{ph}\rangle\nonumber
\end{eqnarray}
one obtains for the first-order contribution
$$
-i \sum_{i=2}^3 \{{1\over 2} \Omega_i |i\rangle\langle
1| + {\rm h.c.}\}\Delta t
$$
and for the second order, omitting terms proportional to $(\Delta t)^2$,
\begin{eqnarray}
& & - \!\! \sum_{i,j=2}^{3}\sum_{{\bf
k}\lambda}
\int_{t_{i-1}}^{t_i}\!\!\! dt'\int_{t_{i-1}}^{t'}\!\!\!
dt''
g_{i{\bf k}\lambda}
g^*_{j{\bf k}\lambda} e^{i(\omega_{i}-\omega_{{\bf k}\lambda})t'}
e^{-i(\omega_{j}-\omega_{{\bf k}\lambda})t''}|i\rangle\langle
j|\nonumber\\
&=& -\!\! \sum_{i,j=2}^{3}\sum_{{\bf
k}\lambda}
\int_{t_{i-1}}^{t_i}\!\!\! dt' e^{i(\omega_i - \omega_j)t'}\int_{0}^{t'-t_{i-
1}}\!\!\! d\tau
 g_{i{\bf k}\lambda} g^*_{j{\bf k}\lambda}
e^{i(\omega_{j}-\omega_{{\bf k}\lambda})\tau}|i\rangle\langle
j|.\label{2.7}
\end{eqnarray}
In the last equation we have substituted $\tau=t'-t''$. Since
$\Delta t$ is much larger than the inverse optical
frequencies
$\omega_{i}^{-1}$ one can extend the inner integral to infinity,
leading to $\pi\delta(\omega_j - \omega_{\bf k\lambda})$ plus a
principle value. Alternatively one can argue that
the correlation function
\begin{equation}
\kappa_{ij}(\tau) = \sum_{{\bf k}\lambda} g_{i{\bf k}\lambda}
g^*_{j{\bf k}\lambda} e^{i(\omega_{j}-\omega_{{\bf k}\lambda})\tau}
\label{2.8}
\end{equation}
is negligible for $\tau\gg \omega_{j}^{-1}$. The sum over ${\bf
k}$ then
yields generalized decay constant $\Gamma_{ij}$ and  level shifts
$\Delta_{ij}$. The level shifts are small \cite{Cohen} and will
be neglected in the following. With
\begin{equation}
\Gamma_{ij} = \frac{e^2{\bf D}_{i1}\cdot{\bf  D}_{1j}}
{6\pi\epsilon_0\hbar c^3} \omega_{j}^3\, \label{2.9a}
\end{equation}
one obtains for the second-order contribution
\begin{equation}
-\sum_{i,j=2}^{3}
\Gamma_{ij}|i\rangle\langle j|
\int_0^{\Delta t} d\tau e^{-i(\omega_i -\omega_j)\tau}\, .\label{2.10}
\end{equation}
Now, if $\omega_3 -\omega_2$ is in the optical range, as supposed
here, then the last integral vanishes for $i \ne j$ and equals
$\Delta t$ otherwise.
We note that
\begin{eqnarray*}
\Gamma_{ii} = \frac{1}{2}\, A_i
\end{eqnarray*}
where $A_i$ is the Einstein coefficient of the $i-$th level.
Collecting all  terms we thus obtain
\begin{equation}
\langle 0_{ph}|U_I(t_i,t_{i-1})|0_{ph}\rangle = {\bf 1}_A
- i \sum_{i=2}^3 \{{1\over 2} \Omega_i |i\rangle\langle
1| + {\rm h.c.}\}\Delta t  -
\sum_{i=2}^{3} {1\over 2} A_i|i\rangle\langle i|\Delta t\,.
\label{2.13}
\end{equation}
This can be written as $\exp\{-i\, H^I_{\rm red}\Delta t/\hbar\}$
where, in matrix
notation, the reduced Hamiltonian  $ H^I_{\rm red}$ and the atomic
operator $M$ are defined through
\begin{equation}
H^I_{\rm red}/\hbar~=  \frac{1}{2}~\left( \begin{array}{ccc}
0 & \Omega_2  & \Omega_3 \nonumber\\
\Omega_2  &  - i\,A_2 & 0 \\
\Omega_3 & 0 &-i\,A_3 \nonumber
\end{array}\right)
{}~\equiv -iM \,.\label{2.12}
\end{equation}
Later on, we will take $A_2 = 0$.
For arbitrary time intervals we thus have, in the interaction
picture,
\begin{equation}
U^I_{\rm red}(t,0) = e^{-iH^I_{\rm red}t/\hbar}
\equiv e^{-Mt}~. \label{2.15}
\end{equation}
The no-photon probability is then, for $t_0=0$ and initial state
$\psi$, or more generally a density matrix $\rho(0)$,
\begin{eqnarray} \label{2.16}
P_0(t;\psi) &=& || e^{-Mt} | \psi \rangle ||^2 \\
P_0(t;\rho(0)) &=& {\rm tr} \left\{ e^{-Mt} \rho(0) e^{-M^{\dagger}t}
\right\} ~. \nonumber
\end{eqnarray}
The probability that the first photon is emitted in $(t,t+dt)$
equals $P_0(t;\psi) - P_0(t +dt;\psi) \equiv w_1 (t;\psi )dt$ where
\begin{eqnarray} \label{2.17}
w_1( t ; \psi) = - \frac{d}{dt}~ P_0(t;\psi)
\end{eqnarray}
is the probability density for the first photon \cite{Norm}.
For small upper level separation nonzero off-diagonal $\Gamma_{ij}$ terms
may appear which lead to interesting coherence effects
\cite{HePl1,HePl2,HePl3,HePl4}. For general $n$-level systems the
reduced Hamiltonian is given in Ref. \cite{He}.

The reduced time development is not unitary. The reason is that
it does not describe the time evolution of the whole ensemble but
that of the sub\-ensemble with no photons. The size of this sub\-ensemble
is decreasing in time since an atom for which a photon has been detected
leaves the sub-ensemble, and this is reflected by the decrease of the norm
squared in Eq. (\ref{2.2a}). The above probability density
determines the (random) time for the first photon. After that the atom
is reset to the ground state, $|1\rangle$,
for a V system. The next emission time
is then determined by $w_1(t;1)$, and so on. In this way one
obtains a quantum trajectory.

 From this description of single systems
one can recover the usual Bloch equations of the complete ensemble as
follows \cite{He}. The density matrix $\rho(t)$ of the ensemble is a
sum of two terms, $\rho^>$ and $\rho^0$, corresponding to a
subensemble of atoms with or without photon emissions until time $t$,
respectively. From Eq. (\ref{2.15}) one has
\begin{equation}\label{2.B1}
\rho^0 ( t ; \rho(0) )  =  e^{- M t} \rho(0)
e^{- M^{\dagger} t}~.
\end{equation}
If $I(\tau;\rho(0))d\tau$ denotes the (unconditioned) probability to
find a photon between $\tau$ and $\tau + d\tau$, then the sub-subensemble
of atoms with their last emission before $t$ in this interval is
described by
\begin{equation}\label{2.B2}
I(\tau;\rho(0))d\tau \rho^0(t - \tau;|1\rangle)
\end{equation}
and therefore
\begin{equation}
\rho^> (t) = ~\int^t_0~d
\tau~I(\tau ; \rho(0) ) \rho^0 (t - \tau;|1\rangle)~.
\label{2.B3}
\end{equation}
Differentiation of $\rho = \rho^0 +\rho^>$ gives
\begin{equation}\label{2.B4}
\dot\rho(t) = \dot\rho^0 ( t ; \rho(0) ) + I(t;\rho(0))|1\rangle
\langle1| + ~\int^t_0~d
\tau~I(\tau ; \rho(0) ) \dot\rho^0 (t - \tau;|1\rangle)~.
\end{equation}
Taking the trace and using tr$\rho(t) \equiv 1$ gives
$I(t ; \rho(0)) = A_2 \rho_{22} + A_3 \rho_{33}$.
 From Eq. (\ref{2.B1}) one obtains $\dot\rho^0$,
and inserting this into Eq. (\ref{2.B4}) gives
\begin{equation} \label{2.B5}
\dot\rho = -\frac{i}{\hbar}[H^I_{\rm red}\rho - \rho
H^{I\dagger}_{\rm red}] + (A_2 \rho_{22} + A_3 \rho_{33})|1\rangle
\langle1|~.
\end{equation}
This is a compact form of the Bloch equations used in Refs.
\cite{B15,B16}.

In this outline of the quantum jump approach state reductions were used as a
tool. But it is noteworthy that one can also use the Markov
approximation, indicating a close connection between the
two  \cite{HeWi}.
\vspace*{1cm}

\noindent {\bf 3. A simple special case: Intermittent probe and $\pi$ pulse.}

\vspace*{1cm}

The quantum jump approach will now be applied to the experimental situation
of Itano et al. \cite{Wine}. Here one can take $A_2 = 0$.
The simplicity of the mechanism becomes particularly clear if the
$\pi$ pulse is switched off while the probe pulse is on. Then
$\Omega_2$ is zero during a probe pulse and Eq. (\ref{2.12}) reads
during this time interval
\begin{equation}\label{0}
H^I_{\rm red}/\hbar ~ = ~\frac{1}{2}~ \left(
\begin{array}{ccc}
0 & 0 & \Omega_3 \\
0 & 0 & 0 \\
\Omega_3 & 0 & - i A_3
\end{array} \right) \, \equiv \; - i M_0~.
\end{equation}
Note that this annihilates the state $| 2 \rangle$ and therefore
the reduced time development leaves $| 2 \rangle$ invariant. At the
end of a probe pulse one has to wait a short transient time of the
order of $A_3^{-1}$ for a possible 3-component to decay. This will
always be done in the following.\\[.5cm]
(i) {\it Effective reduction by a probe pulse} \\[.5cm]
If the state at the beginning of a probe pulse is
\begin{equation}\label{1}
| \psi \rangle = \alpha_1 | 1 \rangle + \alpha_2 | 2\rangle
\end{equation}
and $0 \le \tau \le \tau_{\rm p}$, then $| \psi \rangle$ evolves, until
the emission of the first
photon, as
\begin{equation}\label{2}
e^{- i H^I_{\rm red} \tau / \hbar} | \psi \rangle
= e^{-M_0 \tau } |\psi \rangle
= \alpha_1 e^{- M_0 \tau } | 1 \rangle + \alpha_2 | 2 \rangle~
\end{equation}
since the $\pi$ pulse is assumed to be switched off now.
Due to the term $ A_3 $ in $ M_0 $ the norm of the
first term of the right hand side decreases exponentially and the
first term becomes negligible for $\tau$ large enough. Therefore, if
an atom did not emit a photon until the end of the probe pulse it
will essentially be in the state $| 2 \rangle$, and this happens
with probability given by the norm-squared of the r.h.s., i.e. by
$|\alpha_2|^2$ for large enough $\tau_{\rm p}$.

On the other hand, if an atom does emit one or more
photons -- this happens with
probability $1 - || e^{-M_0 \tau } |\psi \rangle||^2
 = |\alpha_1|^2(1 - || e^{-M_0 \tau } |1 \rangle ||^2)$ --
then right thereafter it is in state $| 1 \rangle$  and
will then be pumped between $|1 \rangle$ and $|3 \rangle$ by
the probe pulse, with photon emissions. A short time after the
end of the probe pulse, $|3 \rangle$ decays to $|1 \rangle$
due to the damping term $A_3$.
Thus a single atom is projected onto
$|1 \rangle$ or $|2 \rangle$ by the probe pulse with probability
$|\alpha_1|^2$ and $|\alpha_2|^2$ if
$|| e^{-M_0 \tau } |1 \rangle ||^2$ can be neglected.
For  an ensemble of atoms
the density matrix becomes diagonal because of the reduction
of every single atomic state.

The preceding analysis is easily made more quantitative as follows. The
eigenvalues of $M_0$  are $\lambda_2 = 0$ and
\begin{equation}\label{3}
\lambda_{1,3} = \frac{1}{4}~ \left( A_3 \pm
\sqrt{A_3^2 - 4 \Omega_3^2} \right)~.
\end{equation}
The first term on the r.h.s. of Eq. (\ref{2}) becomes
\begin{equation}\label{4}
\alpha_1 e^{- M_0 \tau } | 1 \rangle =
\alpha_1~~\frac{1}{\lambda_1 - \lambda_3}
{}~\left\{ (M_0 - \lambda_3 ) e^{- \lambda_1 \tau}
- (M_0 - \lambda_1 ) e^{- \lambda_3 \tau} \right\} | 1 \rangle~,
\end{equation}
as immediately checked by explicit differentiation \cite{gant}.
For $2 \Omega_3 \le A_3$ the root in Eq. (\ref{3}) is real and
$\lambda_1$ and $\lambda_3$ are positive. Therefore
in this case the exponential decrease goes at least as
\begin{eqnarray}\label{5}
\exp \left[ -~\frac{\tau}{4}~\left( A_3 -~\sqrt{A_3^2-4 \Omega_3^2}
\right) \right]
& \le & \exp \left[ -~\frac{\tau}{2} \Omega_3^2 / A_3 \right]
\end{eqnarray}
and this becomes exponentially small for a probe pulse of length
$\tau_{\rm p}$ with
\begin{equation}\label{6}
\tau_{\rm p} \gg 2 A_3 / \Omega_3^2 \hspace*{1cm}
(\mbox{for}~ 2 \Omega_3 < A_3)~.
\end{equation}
If $4 \Omega_3^2 \ge A^2_3$ the root is imaginary and the decrease
goes as
\begin{equation}\label{7}
\exp \left[ -~\frac{\tau}{4}~A_3 \right]
\end{equation}
and Eq. (\ref{6}) is replaced by
\begin{equation}\label{8}
\tau_{\rm p} \gg 4 / A_3 \hspace*{1cm}
(\mbox{for}~ 2 \Omega_3 > A_3)~.
\end{equation}
This can be combined to
\begin{equation}\label{9}
\tau_{\rm p} \gg \max \left\{A_3^{- 1} , A_3 / \Omega_3^2 \right\}~.
\end{equation}
For the special case under consideration this is the
condition on the length of the probe pulse for an effective
reduction to $| 1 \rangle$ and
to $| 2 \rangle$, with probability $| \alpha_1 |^2$ and $|\alpha_2
|^2$, respectively.
\\[.5cm]
(ii) {\em Population vs. observed photons}
\\[.5cm]
A single atom is projected onto the ground state
if it emits several photons during the
probe pulse. For an ensemble of atoms
the number of photons is expected to
be a measure for the population of level
1. With the quantum jump approach
this is easily seen as follows.

The probability for no photon emission until time $\tau,~P_0 (\tau;\psi)$,
is given by the norm squared of the r.h.s. of Eq. (\ref{2}),
according to Eq. (\ref{2.16}), and it approaches $|\alpha_1|^2$ for
large times. For the subensemble of atoms {\em with} emissions
the (conditional) probability density
for the emission of the first photon  is therefore
\begin{equation}
w_1(\tau;\psi)/|\alpha_1|^2
=-\frac{d}{d\tau}P_0(\tau;\psi)/|\alpha_1|^2
\end{equation}
for $\alpha_1 \not= 0$ (for $\alpha_1 = 0$ it is not defined).
Since the two terms on the
r.h.s. of Eq. (\ref{2}) are orthogonal, the $\alpha_2$ term drops out
upon differentiation, and $-d/d\tau~P_0 (\tau;\psi)$ is proportional to
$|\alpha_1|^2$. Thus for $\alpha_1 \not= 0$
the probability density for the first photon  as well as the number
of photons per atom in this subensemble (i.e. the conditional
expectation value) is
independent of the atomic state at the beginning of the probe
pulse. The  number, $N(\tau;\psi)$, of photons per atom for an unconditioned
ensemble with initial state $|\psi\rangle$ (i.e. the usual
expectation value) is
\begin{equation} \label{photnr}
N(\tau;\psi)=|\alpha_1|^2~N(\tau;1)~.
\end{equation}
This expression is now also true for $\alpha_1 = 0$.

The result  is exact for the case $\Omega_2 = 0$
and for all times  $\tau$ within the
validity domain of the quantum jump approach and Bloch equations. As
a consequence, in the case of an ensemble realized by a large number
of noninteracting atoms without cooperative effects, as in Ref.
\cite{Wine}, the number of observed photons per atom is proportional
to the population of level 1.
\\[.5cm]
(iii) {\it Effectiveness of state reduction}
\\[.5cm]
Instead of the condition in Eq. (\ref{9}) for $\tau_{\rm p}$ one can
use   $N(\tau_{\rm p};\psi)$, the number of photons per atom emitted until
time $\tau_{\rm p}$, as a more precise measure for the effectiveness
of state reduction. Eq. (\ref{9}) corresponds to $N(\tau_{\rm p};1) \gg 1$,
but we will show that also for smaller $N(\tau_{\rm p};\psi)$
an almost complete state reduction is obtained.

As pointed out in (i) above, atoms with photon emissions are in
$|1\rangle$ if one waits at the end of the probe pulse  for
a short transient
time to allow for the decay of level $|3\rangle$. But atoms
without emissions, however, still contain a part which is not reduced
to $|2\rangle$
if $\alpha_1 \neq 0$ in Eq. (\ref{2}).
At the end of the probe pulse and  after a short transient
time to allow for the decay of level $|3\rangle$  this non-reduced
component is
\begin{equation}
\alpha_1 ~ \langle 1| e^{-M_0\tau_{\rm p}} |1 \rangle ~ |1 \rangle ~.
\label{nr1}
\end{equation}
The smaller the norm of this, the better the reduction to $|2\rangle$.
The norm can be estimated by
\begin{equation}
\label{nr2}
|| \alpha_1 ~ \langle 1| e^{-M_0\tau_{\rm p}} |1 \rangle ~ |1
\rangle||
\le ||\alpha_1  e^{-M_0\tau_{\rm p}} |1 \rangle || ~.
\end{equation}

For initial state $|1 \rangle$ the number of photons per atom until
time $\tau_{\rm p}$, $N(\tau_{\rm p};1)$,
is in good approximation given by the steady state
emission rate \cite{Loudon} multiplied by $\tau_{\rm p}$,
\begin{equation}
N(\tau_{\rm p} ;1) = A_3 \frac {\Omega_3^2}{A_3^2+2\Omega_3^2}\,
\tau_{\rm p} ~.
\label{N}
\end{equation}
For initial state $|\psi \rangle =  \alpha_1 | 1 \rangle + \alpha_2
| 2\rangle$
we denote the number of photons per atom until time $\tau_{\rm p}$ by $N$,
i.e. $N = N(\tau_{\rm p},\psi)$. Then one
can use Eqs. (\ref{photnr}) and (\ref{N}) to express $\tau_{\rm p}$
through $N$ and $|\alpha_1|^2$,
\begin{equation}
\tau_{\rm p}~=~\frac{A_3^2 + 2\Omega_3^2}{A_3\Omega_3^2}
                            N/|\alpha_1|^2 ~.
\end{equation}
This can now be inserted into Eq. (\ref{nr2})
to obtain an estimate of the non-reduced
part when N photons per atom are emitted. This  norm is easily
calculated by  Eq. (\ref{2}). The norm is a function of
$\Omega_3/A_3$, $N$
and $\alpha_1$, and one easily shows that for fixed other parameters
it becomes largest for
$|\alpha_1| = 1$. A graphical evaluation gives as upper bound for
the norm of the non-reduced part
\begin{equation}
|| \alpha_1 ~ \langle 1| e^{-M_0\tau_{\rm p}} |1 \rangle ~ |1 \rangle ||
{}~\le~1.04\cdot e^{-N(\tau ;1) /2}
\label{bound}
\end{equation}
which holds for $N \ge 2$ and for all values of $\Omega_3$, $A_3$,
and $\alpha_1$. For increasing $N$ the reduction thus becomes very
effective.

For particular values of $\Omega_3/A_3$ the bound for the non-reduced
part can be substantially improved. E.g., for $\Omega_3$ in the vicinity
of $A_3/2$ the non-reduced part becomes much smaller than the above
bound. For very small and very large values of $\Omega_3/A_3$
the reduction is somewhat less efficient than for $\Omega_3$
close to $A_3/2$.

In Table 1 we have listed the norm of the maximally possible
non-reduced part for various values of $N$ and  $\Omega_3/A_3$.
The best reduction occurs for $\Omega_3$ around $A_3/2$,
but the reduction is also excellent for small and large values of
$\Omega_3$ if $N$ is larger than 8.

\begin{table}
\begin{center}
\noindent \begin{tabular}{|c|c|c|c|c|c|c|c|}
\hline
\noindent N & 4 & 5 & 6 & 8 & 10 & 20 & 50 \\
\hline
\noindent $\Omega_3 \ll A_3$
& 0.135 & 0.082 & 0.050 & 0.018 & $6.7 \cdot 10^{-3}$
& $4.5 \cdot 10^{-5}$ & $1.4 \cdot 10^{-11}$  \\
\hline
\noindent $~~\;\Omega_3 = A_3/2$
& 0.023 & 0.006 & 0.002 & 0.0001 & $6.7 \cdot 10^{-6}$
& $4.0 \cdot 10^{-12}$ & $2.9 \cdot 10^{-31}$ \\
\hline
\noindent $\Omega_3 = A_3$
& 0.051 & 0.027 & 0.015 & 0.004 & $6.9 \cdot 10^{-4}$
& $4.3 \cdot 10^{-7}$ & $5.1 \cdot 10^{-17}$ \\
\hline
\noindent $~~\Omega_3 = 2A_3$
& 0.094 & 0.065 & 0.038 & 0.011 & $3.4 \cdot 10^{-3}$
& $1.2 \cdot 10^{-5}$ & $5.7 \cdot 10^{-13}$ \\
\hline
\end{tabular}

\parbox{15cm}{\caption {Maximally possible non-reduced part
for given  number $N$ of observed photons per atom and  for different
values of $\Omega_3/A_3$. The $\pi$ pulse switched off. }}
\end{center}
\end{table}

Summarizing this section, we have shown the following
for the case in which the $\pi$ pulse is turned off during a probe
pulse.

\begin{itemize}
\item The probe pulse provides an effective reduction of
the initial state provided its duration is much longer than $\max
\left\{ 1 / A_3, A_3 / \Omega_3^2 \right\}$, a rather mild
restriction.
\item For   an (infinite) ensemble the observed number
of photons per atom is proportional to the population of level 1.
\item Already for small average numbers of emitted photons
an almost complete state reduction is obtained.
\end{itemize}

\vspace*{1cm}

\noindent {\bf 4. Simultaneous probe and $\pi$ pulse}

\vspace*{1cm}

Now we consider a single probe pulse with the
$\pi$ pulse switched on. At the end of a probe pulse, we include a
short transient time of the order $A_3^{-1}$ to allow for the decay of
the auxiliary level 3. Since this short transient time, with the
action of the $\pi$ pulse, is neglected this introduces an error of
the order $\Omega_2/A_3$ in the time development of the above
subensembles. In the following we
will assume that this error is much smaller than $\epsilon_{\rm p}$.
Since $\epsilon_{\rm p} = A_3\Omega_2/ \Omega_3^2$ this is equivalent
to the condition
\begin{equation}\label{4.0}
\epsilon_{\rm d} \equiv \Omega_3^2/A_3^2  \ll 1~.
\end{equation}
In this and the next section  we will use the conditions
$\epsilon_{\rm p}$, $\epsilon_{\rm d} \ll 1$. The condition
$\epsilon_{\rm R} \ll 1$ is then automatically fulfilled
\cite{general}.

The $\pi$ pulse causes a small additional transition between 1 and 2.
We will show that, as a consequence, an atom with initial state
$| \psi \rangle = \alpha_1 | 1 \rangle + \alpha_2 | 2\rangle$ and
without photon emission
until the end of a probe pulse, including the above short transient
time,
is not in the state $| 2 \rangle$ but
in a state  $|\tilde \lambda \rangle$ which also has a 1-component.
On the other hand, if an atom emits photons, the last photon may have
been emitted some time before the end of the probe pulse. Right after
the emission the atom is in $| 1 \rangle$, but until the end of
the probe pulse a small contribution of state $| 2 \rangle$ may
build up, due to the action of the $\pi$ pulse. Thus the atom will
not be in $|1\rangle$ as in the ideal projection result. Instead it
is in a mixed state, denoted by $\tilde\rho$.
Thus, with the $\pi$ pulse switched on, a single probe pulse effectively
projects onto the state $| \tilde\lambda \rangle$ if no photon is emitted
and onto $\tilde\rho$ otherwise, and this happens with the probability
$P_0(\tau_{\rm p};\psi)$ and $1 - P_0(\tau_{\rm p};\psi)$,
respectively. In the  following
$| \tilde\lambda \rangle$ and $\tilde\rho$
will be  determined.
If $\epsilon_{\rm p},~\epsilon_{\rm d} \ll 1$ the differences
between $| \tilde\lambda \rangle$ and $|2\rangle $, $\tilde\rho$
and $|1 \rangle \langle 1|$ and $P_0(\tau_{\rm p};\psi)$ and
$|\alpha_2|^2$ are  small and the results of
Section 3 can be used as a good approximation.

A single atom now evolves with the reduced Hamiltonian
$H^I_{\rm red} \equiv - i \hbar M$ of Eq. (\ref{2.12}), with $A_2 = 0$,
until the emission of the first photon. The possible
pumping between levels 1 and 2 is reflected by the fact that
$| 2 \rangle$ is no longer annihilated by $H^I_{\rm red}$. To
calculate the time development $\exp \{- i~H^I_{\rm red} t/\hbar\} =
\exp \{- M t\}$ one may proceed in two alternative ways if the
eigenvalues $\lambda_i$ of $M, i = 1, 2, 3,$ are all distinct. In the
first way one determines the corresponding eigenvectors $|
\lambda_i \rangle$ of $M$. Since $M$ is non-hermitian these are in
general nonorthogonal, and therefore one also needs the reciprocal
basis $\{| \lambda^i \rangle\}$ with $ \langle \lambda^i |
\lambda_j \rangle = \delta_{i j}$. Then one can write
\begin{equation}\label{4.1a}
e^{- M t} = \sum~e^{- \lambda_i t} | \lambda_i \rangle \langle
\lambda^i |.
\end{equation}
Alternatively one has, as a generalization of Eq. (\ref{4}),
\begin{equation}\label{4.1}
e^{- M t} = e^{- \lambda_1 t}~~\frac{(M - \lambda_2)(M -
\lambda_3)}{(\lambda_1 - \lambda_2)(\lambda_1 -
\lambda_3)}~~+~\mbox{cyclic permutations}
\end{equation}
which is immediately checked by application to  eigenvectors. The
case of degenerate eigenvalues can be treated by considering appropriate
limits of Eq. (\ref{4.1}).

Comparing the two equations one sees that $| \lambda_i
\rangle \langle \lambda^i |$ coincides with the operator multiplying $e^{-
\lambda_i t}$ in Eq. (\ref{4.1}). Moreover, applying this operator  to
any vector gives a multiple of $| \lambda_i \rangle$, thus automatically
yielding the eigenvectors. The eigenvalues are the roots of the
characteristic equation \cite{char} which, in principle, can be solved
in closed form. One easily calculates
\begin{eqnarray}\label{4.5}
\left( M - \lambda_1 \right) \left( M - \lambda_3 \right)
\left( \begin{array}{c}
    1 \\ 0 \\ 0
\end{array} \right)  & = & \left( \begin{array}{c}
  \lambda_2 (\lambda_2 - ~\frac{1}{2}~A_3)\\
  \frac{1}{2}~i\Omega_2 (\lambda_2 -~\frac{1}{2}~A_3) \\
  \frac{1}{2}~i\lambda_2 \Omega_3
  \end{array} \right) \nonumber~.
\end{eqnarray}
By the above remarks, this is a multiple of $|\lambda_2 \rangle$.
Similarly \cite{id},
\begin{equation}\label{4.4}
(M - \lambda_1) (M - \lambda_3) ~
\left( \begin{array}{c}
    0 \\ 1 \\ 0
\end{array} \right) = - ~\frac{\Omega_2}{2i \lambda_2}
   ~(M - \lambda_1)(M- \lambda_3) ~
\left( \begin{array}{c}
    1 \\ 0 \\ 0
\end{array} \right)~,
\end{equation}
which is also a multiple of $|\lambda_2 \rangle$.

For the parameter range of the Introduction good approximations for
$\lambda_i$ are
\begin{eqnarray}\label{4.3}
\lambda_{1,3} & = & \frac{1}{4}~ \left( A_3 \pm~\sqrt{A^2_3 - 4
\Omega_3^2} \right) \nonumber \\
\lambda_2 & = & \frac{1}{2}~ A_3 \Omega_2^2/\Omega_3^2
= \frac{1}{2} \epsilon^2_{\rm R} A_3
\end{eqnarray}
where $\lambda_2$ has been obtained by Newton's method. Note that, for
$\Omega_2 \ll \Omega_3,$
\begin{equation}\label{4.3a}
\lambda_2 \ll {\rm Re}~\lambda_{1,3}~.
\end{equation}
Hence the exponentials $\exp \left\{ - \lambda_{1,3} t \right\}$
in Eq. (\ref{4.1}) drop off very rapidly. When calculating $\exp \{ -M
\tau_{\rm p}\} |\psi \rangle$ by Eq. (\ref{4.1}) one can therefore,
as in Section 3, neglect the $|\lambda_1 \rangle$ and $|\lambda_3
\rangle$ terms if $\tau_{\rm p} \gg $ max$\{A^{-1}_3, A_3 /
\Omega^2_3 \} = A_3/\Omega_3^2$ (because $\epsilon_{\rm d} \ll 1$). \\[.5cm]
(i) {\it Subensemble without photon emission} \\[.5cm]
With Eqs. (\ref{4.1}) to (\ref{4.3a}) one can now obtain the state of
the subensemble of
atoms without photon emissions until the end \cite{end} of the probe
pulse. For initial state $|\psi \rangle$,
$$|\psi \rangle = \alpha_1 | 1 \rangle + \alpha_2 | 2 \rangle, $$
it is described at time $\tau_{\rm p}$ by
\begin{eqnarray}\label{4.7}
e^{- M \tau_{\rm p}}|\psi\rangle  &=&
e^{- M \tau_{\rm p}} \left\{ \alpha_1 | 1 \rangle + \alpha_2 | 2
\rangle \right\} \nonumber \\
& = & e^{- \lambda_2 \tau_{\rm p}} \left( \alpha_2 - i \epsilon_{\rm p}
\alpha_1 \right) \sqrt{1 - \epsilon^2_{\rm R} + \epsilon^2_{\rm p}} ~|
\lambda_2 \rangle
\end{eqnarray}
where
\begin{equation}\label{4.7a}
| \lambda_2 \rangle = ~\frac{1}{\sqrt{1 - \epsilon^2_{\rm R} +
\epsilon^2_{\rm p}}}~~\left( \begin{array}{c}
                   - i \epsilon_{\rm p} \\
                   1 - \epsilon^2_{\rm R} \\
                   \epsilon_{\rm R}
                          \end{array} \right) ~.
\end{equation}
Moreover,
\begin{equation}\label{4.7b}
\lambda_2 \tau_{\rm p} =~\frac{1}{2}\epsilon_{\rm p} \pi \frac{\tau_{\rm
p}}{T} \ll 1
\end{equation}
and hence $\exp \left\{ - \lambda_2 \tau_{\rm p} \right\} \approx 1.$
Similarly for an initial density matrix.
The state $|\lambda_2 \rangle$ in Eq. (\ref{4.7a}) has a very small third
component, and after the probe pulse has been turned off this component
will, on the time scale of $A_3^{-1}$, decay to zero \cite{photon}.
Thus at the end of a probe pulse and after this transient decay time
the subensemble with no photons is in the -- normalized -- state
\begin{equation}\label{4.7c}
| \tilde{\lambda} \rangle ~\equiv~ I\!\!P_{1,2} | \lambda_2 \rangle
/ \| \cdot \| ~=~\frac{1}{\sqrt{1 + \epsilon^2_{\rm p}}}~~
\left( \begin{array}{c}
                       - i \epsilon_{\rm p}\\
                       1 \\
                       0 \end{array} \right) \approx |2 \rangle
\end{equation}
where $I\!\!P_{1,2} \equiv |1 \rangle \langle 1 | + |2 \rangle
\langle 2 |$  denotes the projector onto the 1-2 subspace and where
terms of higher orders in $\epsilon_{\rm p}$ and $\epsilon_{\rm d}$
have been omitted. The  probability
for no photon emission is
\begin{eqnarray}
P_0(\tau_{\rm p};\psi) &=&
|| I\!\!P_{1,2} e^{- M \tau_{\rm p}}  |\psi \rangle ||^2  \nonumber\\
                       &=&
|\alpha_2 |^2 + 2{\rm Im}~ \alpha_1 \alpha_2^* \epsilon_{\rm p} -
\pi |\alpha_2 |^2 \frac{\tau_{\rm p}}{T_\pi} \epsilon_{\rm p} ~~~~+~
\mbox{ higher orders in } \epsilon_{\rm p},~\epsilon_{\rm d}.
\label{4.7d}
\end{eqnarray}
For an initial density matrix $\rho^{\rm in}$ instead of the pure state
$\psi$ one has to replace $|\alpha_2 |^2$  by $\rho^{\rm in}_{22}$ and
$\alpha_1 \alpha_2^*$ by $\rho^{\rm in}_{12}$.

Thus, to good approximation, the probability for no photon
emission is proportional to the population of level 2 and the atoms
with no emissions are approximately in the state $|2 \rangle$.
For the parameters \cite{remark} of the experiment \cite{Wine}
the corrections in Eq. (\ref{4.7d}) are less than $4 \cdot 10^{-4}$.
\\[0.5cm]
(ii) {\it Subensemble} with {\it photon emissions}\\[0.5cm]
We will now calculate the density matrix $\tilde \rho$ for the
subensemble with  photon emissions. One can employ a
systematic expansion in powers of $\Omega_2$, including second order.
However, the following
more physically motivated procedure is simpler and yields the same
results.

At the beginning of the probe pulse we assume the ensemble  to be in
the pure state $| \psi \rangle$; for
a density matrix the treatment is similar. The
complete ensemble at
time $\tau_{\rm p}$ after the  beginning of the
probe pulse
can be thought of as consisting of two subensembles
of atoms with and without photon emissions.
The latter is described by Eq.  (\ref{2.B1}),
\begin{equation}\label{4.9a}
\rho^0 ( \tau_{\rm p} ; \psi )  \equiv  e^{- M \tau_{\rm p}} | \psi
\rangle \langle \psi |  e^{- M^{\dagger} \tau_{\rm p}}~,
\end{equation}
with its relative weight  given by Eq. (\ref{2.16}),
\begin{equation}\label{4.9aa}
{\rm tr} \rho^0(\tau_{\rm p};\psi)~=~ P_0(\tau_{\rm p};\psi)~.
\end{equation}
According to Eq. (\ref{2.B3})
the former subensemble is described by
\begin{equation}
\rho^> (\tau_{\rm p} ; \psi ) = ~\int^{\tau_{\rm p}}_0~d
\tau~I(\tau ; \psi ) \rho^0 (\tau_{\rm p} - \tau;1)~.
\label{4.10}
\end{equation}
The complete density matrix is $\rho^> + \rho^0$, and therefore
\begin{equation}\label{4.9ab}
{\rm tr} \rho^> ~=~ 1 -P_0~.
\end{equation}

We now determine  $\rho^>_{12}$ and $\rho^>_{22}$ at time $\tau_{\rm
p}$.
For initial state $|1\rangle$ the no-photon probability
decreases rapidly, and therefore
$\rho^0 (\tau_{\rm p}-\tau;1)$ contributes essentially  only in the
vicinity of $\tau \approx \tau_{\rm p}$ (cf. Eqs. (\ref{4}) to
(\ref{6})). Because of this one can replace
$I(\tau;\psi)$ by $I(\tau_{\rm p};\psi)$ which is approximately
equal to the fraction $1 - P_0(\tau_{\rm p};\psi)$
of emitting atoms times $I(\tau_{\rm p};1)$.
The latter practically equals the stationary emission rate
for the three-level system which is, up to terms of order
in $\Omega_2^2$, the stationary rate from Eq.
(\ref{N}) for the two-level system.
For the calculation of the 12 and 22 component these
corrections in $\Omega_2^2$ can be omitted since
$\rho^0 (\tau_{\rm p}-\tau;1)_{12}$ and
$\rho^0 (\tau_{\rm p}-\tau;1)_{22}$ are themselves of order $\Omega_2$
and $\Omega_2^2$ and much smaller than $1$. Thus we obtain
\begin{equation} \label{int3}
\rho^> (\tau_{\rm p} ; \psi )_{12/22}
= (1 - P_0(\tau_{\rm p};\psi)) \frac{A_3 \Omega_3^2}{A_3^2+2\Omega_3^2}
\int^{\tau_{\rm p}}_0~ d\tau \rho^0(\tau;1)_{12/22}~.
\end{equation}

After the end of the probe pulse, any population of level 3 will
rapidly decay to level 1 in a transient time of order $A_3^{-1}$. We
denote the resulting
-- normalized -- density matrix of the subensemble of atoms with
emissions by $\tilde{\rho}$  \cite{photon2}. The normalization factor of
$\rho^>$ is $1 - P_0$, and  according to Eq. (\ref{4.9ab}) we obtain
from Eq. (\ref{int3})
\begin{equation} \label{int4}
\tilde{\rho} (\tau_{\rm p} ; \psi )_{12/22}
=  \frac{A_3 \Omega_3^2}{A_3^2+2\Omega_3^2}
\int^{\tau_{\rm p}}_0~ d\tau \rho^0(\tau;1)_{12/22}~.
\end{equation}
For this we note that, by Eq. (\ref{4.9a}),
\begin{eqnarray}
 \rho^0_{12}(\tau;1) & = & \langle 1| e^{-M \tau} |1 \rangle
 \langle 1| e^{-M \tau} |2 \rangle  \nonumber\\
 \rho^0_{22}(\tau;1) & = & | \langle 2| e^{-M \tau} |1 \rangle |^2~.
\label{int6}
\end{eqnarray}
Using Eq. (\ref{4.1}) for $e^{-M\tau}$ an elementary
calculation yields
\begin{eqnarray}
  \tilde{\rho}_{22} &=&  \pi
  \frac{\tau_{\rm p}}{T_\pi}~\epsilon_{\rm p}~~~~
 +~ \mbox{higher orders in } \epsilon_{\rm p},~\epsilon_{\rm d}
\label{4.12d}\\
  \tilde{\rho}_{12} &=& i  \epsilon_{\rm p} ~~~~~~~~ +~
\mbox{higher orders in } \epsilon_{\rm p},~\epsilon_{\rm d}
\label{4.12e}\\
  ~ \tilde{\rho}_{11} &=& 1-\tilde{\rho}_{22}~,~~~
  \tilde{\rho}_{21} ~=~ -\tilde{\rho}_{12} ~
\end{eqnarray}
and its $13$, $23$ and $33$  components  vanish,
\begin{equation}
 ~ \tilde{\rho}_{13} ~=~\tilde{\rho}_{31}
{}~=~\tilde{\rho}_{23} ~=~\tilde{\rho}_{32}
{}~=~\tilde{\rho}_{33} ~=~0. ~
\end{equation}

We conclude that after a short transient time at the end of the probe pulse
the subensemble {\em with} photon emissions is described by the above
normalized state $\tilde{\rho}$, which is independent of the initial
state $|\psi \rangle$, and the relative size of the subensemble is given
by $1 -P_0(\tau_{\rm p};\psi)$.

We note that $\tilde{\rho}_{22}$ and $\tilde{\rho}_{12}$ are indeed very
small. For the parameters of the experiment \cite{Wine,remark} one has
$\tilde{\rho}_{22} < 1.2 \cdot 10^{-5}$ and
$|\tilde{\rho}_{12}| < 4.1 \cdot 10^{-4}$. \\[0.5cm]
(iii) {\em Level population after a probe pulse} \\[0.5cm]
We denote by  $\rho^{\rm (p)}$ the density matrix of the complete
atomic ensemble after a short transient time at the end of a probe
pulse. By the preceding results it is given by
\begin{equation} \rho^{\rm (p)} = P_0(\tau_{\rm p};\rho^{\rm in})
|\tilde\lambda\rangle
\langle\tilde\lambda| + (1 - P_0(\tau_{\rm p};\rho^{\rm in})) \tilde\rho
\label{4.12f}
\end{equation}
where $\rho^{\rm in}$ is the density matrix at the beginning of the probe
pulse. From Eqs. (\ref{4.7d}), (\ref{4.7c}) and (\ref{4.12d})
one immediately obtains
for the population of level 2 after the probe pulse
\begin{equation} \label{4.12g}
\rho_{22}^{\rm (p)} = \rho^{\rm in}_{22} + 2\epsilon_{\rm p}~
{\rm Im}\rho^{\rm in}_{12}
+ \pi \frac {\tau_{\rm p}}{T_\pi}\epsilon_{\rm p}
(1 - 2 \rho^{\rm in}_{22})~.
\end {equation}
The first term is the projection-postulate result for an ideal
measurement. For $n$ probe pulses, $\rho_{12}$ is of the order
$\sin \frac{\pi}{2}
\left( \frac{1}{n}  - \frac{\tau_{\rm p}}{T_{\rm \pi}} \right)$, and
if $n$ is as in the experiment \cite{Wine}  the second term is
larger than, or comparable to, the last term. In the corresponding
Eq. (16) of Ref. \cite{B16} this important term is missing, due the
approximation used there \cite{corr}, and the correction appearing
there is equivalent to $O(\epsilon_{\rm p}2\pi \tau_{\rm p}/T_{\pi})$
in our notation.\\[0.5cm]

Summarizing this section, we have shown that, for the parameters
$\epsilon_{\rm p}$ and   $ \epsilon_{\rm d}$
much less than 1, a probe pulse acts as an effective state reduction also
in the presence of the $\pi$ pulse. The reductions are to
$\tilde{\rho}$ and $| \tilde{\lambda} \rangle$,
corresponding to subensembles with and without emissions,
respectively. One has  $\tilde{\rho} \approx |1\rangle \langle
1|$ and $|\tilde{\lambda}\rangle \approx |2 \rangle$.
The corrections have been explicitly calculated in terms of the above
parameters \cite{rem3}.

\vspace*{1cm}

\noindent {\bf 5. Level population after n probe pulses}

\vspace*{1cm}

The preceding results show that, after a probe pulse and a short transient
time, the ensemble consists of two subensembles, one in the state
$\tilde \rho $ and the other in $| \tilde{\lambda} \rangle$,
corresponding to atoms with and without photon emissions. The small
difference from the projection result can have a cumulative effect
for the density matrix after $n$ probe pulses.

After a probe pulse, the density matrix for the complete ensemble is of
the form
\begin{equation}\label{5.1}
\rho = \alpha ~ \tilde \rho + \beta ~
|\tilde{\lambda} \rangle \langle \tilde{\lambda} |
\end{equation}
with $\alpha + \beta = 1$.  Until the beginning of the next probe pulse
at a time $T_{\rm \pi}/n - \tau_{\rm p}$ later
the time development is given by the $\pi$ pulse only, i.e. in matrix
notation and in the $|1 \rangle - |2 \rangle$ subspace by
\begin{equation}\label{1.1}
 U_\pi (t, 0) = \left( \begin{array}{cc}
\cos~\frac{1}{2}~\Omega_2 t & - i \sin~\frac{1}{2}~\Omega_2 t\\
- i \sin~\frac{1}{2}~\Omega_2 t & \cos~\frac{1}{2}~\Omega_2 t
                             \end{array} \right).
\end{equation}
For $t = \frac{T_ \pi}{n} - \tau_{\rm p}$ we define $U_\pi (t,0) \equiv
\tilde{U}_n$.  We will now determine the density matrix after the
$k$-th probe pulse. To this end we put
\begin{eqnarray}\label{5.2}
p & = & {\rm tr}   \left\{ I\!\!P_{1,2}
e^{-M  \tau_{\rm p}}   \tilde{U}_n
\tilde\rho   \tilde{U}^{\dagger}_n
e^{-M^{\dagger}\tau_{\rm p}}   I\!\!P_{1,2}
\right\}  \nonumber\\
q & = & ||   I\!\!P_{1,2}   e^{-M \tau_{\rm p}}   \tilde{U}_n
|\tilde{\lambda} \rangle   ||^2~,
\end{eqnarray}
where $I\!\!P_{1,2}$ is the projector onto the $1-2$ subspace. Physically,
$p$ is the probability of finding no photons after the next probe pulse
\cite{end} if one had started with $\tilde\rho$ at the end of the preceding
probe
pulse. Similarly for $q$ and $| \tilde{\lambda} \rangle$.

With the abbreviation
\begin{eqnarray}\label{5.2a}
s_n & = & \sin  \pi  \left( \frac{1}{n}  -
\frac{\tau_{\rm p}}{T_{\rm \pi}} \right) \nonumber\\
c_n & = & \cos  \pi  \left( \frac{1}{n}  -
\frac{\tau_{\rm p}}{T_{\rm \pi}} \right)
\end{eqnarray}
one finds by a straightforward calculation from Eq. (\ref{1.1}) for
$U_\pi$ and Eq. (\ref{4.1}) for $e^{- M \tau_{\rm p}}$
\begin{eqnarray}\label{5.2b}
p & = & \frac{1}{2} (1-c_n) + s_n \epsilon_{\rm p}
 +\tilde\rho_{22}  c_n
 -i\tilde\rho_{12} s_n
 -\frac{1}{2} \pi \frac{\tau_{\rm p}}{T_{\pi}} (1-c_n)~ \epsilon_{\rm p}
\nonumber\\
& = & \frac{1}{2} (1-c_n) + 2 s_n  \epsilon_{\rm p}
 + \pi\frac{\tau_{\rm p}}{T_{\pi}} c_n \epsilon_{\rm p}
 - \frac{1}{2} \pi \frac{\tau_{\rm p}}{T_{\pi}} (1-c_n)~ \epsilon_{\rm p}
\nonumber \\
q & = & \frac{1}{2} (1+c_n) - 2 s_n \epsilon_{\rm p}
 - \frac{1}{2} \pi \frac{\tau_{\rm p}}{T_\pi}  (1+c_n)~ \epsilon_{\rm p}
\end{eqnarray}
where  higher orders in $\epsilon_{\rm p}$ and $\epsilon_{\rm d} $
have been omitted.

Now suppose that after the $(k - 1)$-st probe pulse the density matrix
$\rho$ is given by Eq. (\ref{5.1}) with $\alpha = \alpha(k-1)$ and
$\beta = \beta (k - 1)$. Then after the $k$-th  probe pulse the
relative size of the no-photon subensemble is given by
\begin{eqnarray}\label{5.3}
\beta (k) & = & p~\alpha(k - 1) + q ~\beta (k
- 1) \nonumber\\
& = & p + (q - p) ~\beta (k - 1)
\end{eqnarray}
where $\alpha = 1 - \beta$ has been used. The solution of this
recurrence relation is
\begin{eqnarray}\label{5.4}
\beta (k) & = & p~\frac{1-(q-p)^{k-1}}{1-(q-p)} +
(q-p)^{k-1} ~ \beta (1) \nonumber\\
\alpha (k) & = & 1 - \beta (k)~.
\end{eqnarray}
According to Eq. (\ref{5.1}) $\beta (1)$ is the probability to find
no photon during the first probe pulse \cite{end}.
For the initial condition
that all atoms are prepared in the ground state at the beginning of the
experiment, as in the experiment of Ref.  \cite{Wine}, $\beta(1)$
is given by
\begin{eqnarray}
\beta (1) &=&
|| I\!\!P_{1,2} e^{-M \tau_{\rm p}} \tilde{U}_n |1 \rangle ||^2
\nonumber \\
&=& \frac{1}{2} (1-c_n) + s_n ~ \epsilon_{\rm p}
- \frac{1}{2} \pi\frac{\tau_{\rm p}}{T_{\pi}} (1-c_n) ~ \epsilon_{\rm p} ~.
\label{beta}
\end{eqnarray}

At the end of the $\pi$ pulse, i.e. after  the $n$-th probe pulse
the density matrix is
\begin{equation}\label{5.5}
\rho (T_{\rm \pi})  =  \alpha(n) ~\tilde\rho + \beta
(n) ~| \tilde{\lambda} \rangle \langle \tilde{\lambda} |~.
\end{equation}
The populations of levels 2 is then
\begin{eqnarray}\label{5.55}
\rho_{22} (T_{\rm \pi}) & \equiv &
\langle 2 | \rho (T_{\rm \pi}) | 2 \rangle = \alpha(n) ~\tilde\rho_{22}
+ \beta (n) ~| \langle 2 | \tilde{\lambda} \rangle |^2 ~.
\end{eqnarray}
Using Eqs. (\ref{4.7c}), (\ref{4.12d}) and Eqs. (\ref{5.2b})-(\ref{5.55})
and omitting higher orders in $\epsilon_{\rm p}$ and $\epsilon_{\rm d} $
one obtains
\begin{eqnarray}\label{5.6}
\rho_{22} (T_{\rm \pi}) & = & \frac{1}{2} (1-c_n^n)
+ (2n-1) s_n c_n^{n-1}  \epsilon_{\rm p}
+\pi n \frac{\tau_{\rm p}}{T_{\rm \pi}} c_n^n  \epsilon_{\rm p}
\nonumber \\
\rho_{11} (T_{\rm \pi}) & = & 1- \rho_{22} (T_{\rm \pi}) ~.
\end{eqnarray}
Omitting also the
$\epsilon_{\rm p}$ terms one obtains for the populations of
levels 1 and 2 the approximate results \cite{rem4}
\begin{eqnarray}\label{5.8}
\rho_{22} (T_{\rm \pi}) & \cong & \frac{1}{2}(1 - c^n_n) =
\frac{1}{2} \left[ 1 - \cos^n \pi \left(
\frac{1}{n} - \frac{\tau_{\rm p}}{T_{\rm \pi}} \right) \right]
\nonumber\\
\rho_{11} (T_{\rm \pi}) & \cong & \frac{1}{2}(1 + c^n_n) =
\frac{1}{2} \left[ 1 + \cos^n \pi \left(
\frac{1}{n} - \frac{\tau_{\rm p}}{T_{\rm \pi}} \right) \right] ~.
\end{eqnarray}

For $n$ ideal measurements with the projection postulate the result
would be \cite{Wine,B2}
\begin{eqnarray}\label{1.9}
\rho_{22} (T_{\rm \pi}) =~\frac{1}{2}~ \left[ 1 - \cos^n\frac{\pi}{n}
\right]\nonumber\\
\rho_{11} (T_{\rm \pi}) =~\frac{1}{2}~ \left[ 1 + \cos^n\frac{\pi}{n}
\right]~.
\end{eqnarray}
This differs from the result in Eq. (\ref{5.8}) only by the term
$\tau_{\rm p}/T_{\rm \pi}$ in the cosine, and therefore Eq. (\ref{5.8})
can be obtained from the ideal projection result by neglecting
the action of the $\pi$ pulse during a probe pulse,
thus replacing the
time $\Delta T = T_\pi/n$ between measurements by the effective time
$\Delta T = T_\pi/n~ - \tau_{\rm p}$.
This can be also
understood directly quite easily as follows. Atoms which emit
photons during a probe pulse flip repeatedly between levels 1 and
3, and so the $\pi$ pulse acts less effectively them. Moreover, right after
emission of a photon an atom is in the ground state and since the action of
the $\pi$ pulse is of cosine form it  is small for small times.
Similarly, as shown in Section 4, the atoms without emissions  rapidly
approach the state $|\tilde \lambda \rangle \cong |2\rangle$
so that one again has a small action for small times.
Therefore the action of the $\pi$ pulse is greatly inhibited during a
probe pulse .

It is evident from Eq. (\ref{5.8}) that this approximation yields the
same result as if one had switched off the $\pi$ pulse during the
probe pulse and then uses the results of Section 3.

The corrections of Eq. (\ref{5.6}) to the
approximate values for $\rho_{ii} (T_{\rm \pi})$ in Eq. (\ref{5.8}) are
small for $\epsilon_{\rm p} \ll 1$. Moreover, it is
straightforward to show that in the parameter range considered here
the correction to $\rho_{22}(T_{\rm \pi})$ is positive and
increases with $n$, as long as $\tau_{\rm p}$ is not too close to
$T_\pi/n$.  This is borne out in Table 2 where predicted and
observed values of $\rho_{22} (T_{\rm \pi})$ for the parameters of the
experiment of Ref. \cite{Wine} are shown. The second column is based on the
projection postulate for $n$ ideal measurements. The third column is
based on Eq. (\ref{5.8}) or, alternatively, on $n$ ideal
measurements with ensuing switch-off of the $\pi$ pulse for $\tau_{\rm p}$
seconds. The agreement between the quantum jump result in Eq.
(\ref{5.6}) and
the  numerical solution of the three-level
Bloch equations of Eq. (\ref{2.B5}) in column 5
is apparent. The projection postulate with $\Delta T$ modified to
$\Delta T = T_\pi /n - \tau_{\rm p}$  also gives very good results.
The experiment deals with a system where additional energy levels may
make minor contributions \cite{compl} and this may  explain the
deviations from the  experimental  results in the last column of Table 2.

\begin{table}
\begin{center}
\noindent \begin{tabular}{rccccc}
\hline
\noindent & $~~~~~~~\;$Project & \hspace{-1.45cm}ion Postulate  \\
\noindent $n$
& $\Delta T = T_\pi/n$
& $\Delta T = T_\pi/n-\tau_{\rm p}$
& Quantum Jump
& Bloch equation.
& Observed \\
\hline
1 & 1.00000 & 0.99978 & 0.99978 & 0.99978 & 0.995 \\
2 & 0.50000 & 0.49957 & 0.49960 & 0.49960 & 0.500 \\
4 & 0.37500 & 0.35985 & 0.36062 & 0.36056 & 0.335 \\
8 & 0.23460 & 0.20857 & 0.20998 & 0.20993 & 0.194 \\
16 & 0.13343 & 0.10029 & 0.10215 & 0.10212 & 0.103 \\
32 & 0.07156 & 0.03642 & 0.03841 & 0.03840 & 0.013 \\
64 & 0.00371 & 0.00613 & 0.00789 & 0.00789 & $\!\!$-0.006 \\
\hline
\end{tabular}
\parbox{15cm}{\caption { Predicted and observed population
of level 2 at the end
of the $\pi$ pulse for $n$ probe pulses of length $\tau_{\rm p}$.}}
\end{center}
\end{table}

\vspace*{1cm}

\noindent {\bf 6. Conclusions. Does the Zeno effect exist?}

\vspace*{1cm}

We have investigated the so-called quantum Zeno effect for
an ensemble of atomic three-level systems as that of the experiment
in Ref. \cite{Wine}. There has been some controversy about the
interpretation of that experiment as to whether it provides an
experimental proof of that effect. The Zeno effect is a theoretical
prediction for the behavior of a system under rapidly repeated
measurements with ensuing state reductions according to the
projection postulate. As explained in the Introduction, the
controversy has mainly focused on two points, namely (i) whether
some of the ``measurements" in Ref. \cite{Wine} should not rather be
included as ``interactions" in the Hamiltonian, and (ii) whether the
projection postulate is appropriate at all.

Now, a measurement on a microscopic system is indeed a complicated
interaction with another system, ultimately at the macroscopic level.
Not all conceivable measurements conform to the idealized
case considered by von Neumann and L\"uders \cite{N} where each
measurement is associated
with an operator, $A$ say, such that the mean value, $\langle A
\rangle$, is given by the expectation value and the mean-square
deviation by the expectation of $(A - \langle A \rangle)^2$. As a
simple example one may consider a system of photons in a cavity. The
outcome of the measurement consists of two possible pointer readings,
$0$ and $1$ say. The actual measurement is performed by a ``black
box" which has been so constructed that an auxiliary two-level atom
in some initial state is passed through the cavity and then it is
determined whether the atom is in the ground or excited state,
yielding the pointer readings $0$ or $1$, respectively. Considered
as a measurement performed by the black box on the photon system, it
is obviously not of the above kind. There is indeed an operator $A$
for the photon system, whose expectation value gives the mean, but it
has eigenvalues different from $0$ and $1$, and right after a
measurement the photon is not in an eigenstate of $A$. As a
consequence, the mean-square deviation is not of the above form. By
combining part of the black box, namely the atom, with the photons to
a larger system one may possibly retrieve or come close to the
situation considered by von Neumann, depending on how the measurement
on the atom is actually performed.

This example shows that one may have to study the measurement at hand
more closely. This is what we have done with the atomic Zeno
experiment by means of the quantum jump approach \cite{HeWi,He},
which is essentially equivalent to the Monte-Carlo wave function
approach \cite{MC}  and to the quantum trajectory approach \cite{QT}. The
advantage of this approach is that it allows a physically intuitive
as well as analytic treatment of the problem.
Part of the measurement interaction -- the laser pulses -- has
been incorporated in the Hamiltonian; this corresponds in the above
example to combining the photon system and atom to a larger system
and is related to shifting the Heisenberg ``cut" \cite{cut}.

Our analysis has shown that, to a fair accuracy and within the
parameter regimes considered in this paper, a short probe
pulse can indeed be viewed as performing a measurement of levels $1$
or $2$ of the atom, with ensuing state reduction, as given by the
projection postulate for an ideal measurement.
However, since these ``measuring" pulses have
been modeled very accurately in the Hamiltonian we were able to show
that the more realistic case is also slightly more complicated,
giving rise to corrections to the idealized case. These corrections
were explicitly calculated, and they have a cumulative effect on the
density matrix when the
number of probe pulses is increased -- i.e. just for the
interesting case.

What then remains of the Zeno effect? Does it exist at all? In our
opinion  the answer depends on one's point of view. If one takes the
view that for  example the probe  pulses (``measuring pulses")
have nothing to
do with a measurement but just lead to additional terms in the
Hamiltonian, then any change in the temporal development is not
surprising and may simply be attributed to these additional
interaction terms. The other -- possibly more fruitful -- point of
view is that these pulses approximately realize measurements with
state reductions, and then one immediately has simple predictions for
the approximate behavior of the system and arrives at the impediment
and slow-down of the time evolution without complicated
calculation. Finer details require of
course a finer analysis, as performed in the previous sections. An
actual freezing of the state does not seem possible since all
realistic measurements take a finite time. In the present case this
hinges on the corrections and on the finite duration of the probe pulse
(including the transient decay time of order $A_3^{-1}$).

Our analysis may possibly shed some light on the use of the
projection postulate in quantum optics in general, not only in
connection with the Zeno effect. It seems that quite often the
projection postulate is a useful tool  which can give quick and
fairly accurate answers. The accuracy depends on how far the
particular realistic measurement differs from an ideal measurement as
considered in orthodox quantum mechanics, and corrections may have to
be taken into account. The idealization of realistic measurements and
the projection postulate may often be very useful. Over-idealization,
however, should be avoided since any idea, when carried to extremes,
easily reduces itself to absurdity.

\vspace{.5cm}

{\bf Acknowledgments}. One of us (G.C.H) is indebted to the late
Gerhart L\"uders (1920 - 1995),
who also discovered the TCP Theorem, for stimulating discussions on
his formulation of the projection postulate.

\newpage

\begin{figure}
\unitlength 0.6cm
\begin{picture}(18,11)
\thicklines
\put(6,9.5) {\line(1,0){3}}
\put(10,3) {\line(1,0){3}}
\put(7.5,1.5) {\line(1,0){4}}
\thinlines
\put(4.5,3.5) {\line(1,0){0.5}}
\put(5.5,3.5) {\line(1,0){0.5}}
\put(5,5.5) {\line(1,0){0.5}}
\put(5,3.5) {\line(0,1){2}}
\put(5.5,3.5) {\line(0,1){2}}
\put(7,9.5) {\vector(1,-4){2}}
\put(7.5,9.5) {\vector(1,-4){2}}
\put(9,1.5) {\vector(-1,4){2}}
\put(10,1.5) {\vector(1,1){1.5}}
\put(11.5,3) {\vector(-1,-1){1.5}}
\put (5.4,9.5){3}
\put (6,4.25){$\Longrightarrow $}
\put (6.75,6){$\Omega_3$}
\put (8.7,6){$A_3$}
\put (4,2.75){probe pulse}
\put (13.25,3){2}
\put (12.75,2){$\Longleftarrow $}
\put (14.25,2){$\pi$ pulse, $\Omega_2$}
\put (11.75,1.5){1}
\end{picture}

{\caption{V system with (meta-) stable level 2 and Einstein
coefficient $A_3$ for level 3. $\Omega_2$ and $\Omega_3$ are the Rabi
frequencies of the two lasers.}}
\end{figure}
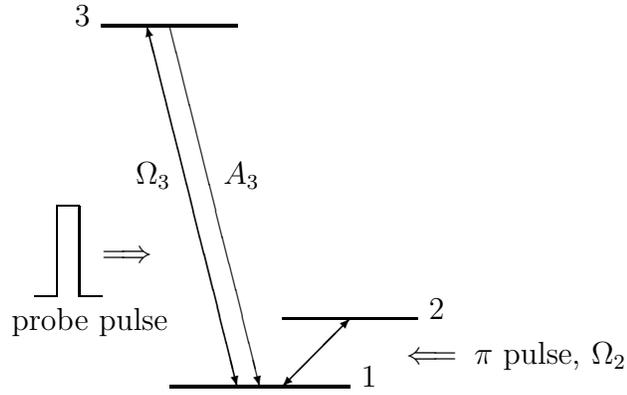

\begin{figure}
\unitlength 0.9cm
\begin{picture}(15.5,6.5)
\thicklines
\put (2,1) {\vector (1,0) {12}}
\put (2,1) {\vector (0,1) {4.5}}

\thinlines
\put (3.5,4.5) {\line (1,0){0.5}}
\put (5.5,4.5) {\line (1,0){0.5}}
\put (7.5,4.5) {\line (1,0){0.5}}
\put (11,4.5) {\line (1,0){0.5}}
\put (2,1.5) {\line (1,0){7}}
\put (10,1.5) {\line (1,0){1.5}}
\put (3.5,1) {\line (0,1){3.5}}
\put (5.5,1) {\line (0,1){3.5}}
\put (7.5,1) {\line (0,1){3.5}}
\put (11,1) {\line (0,1){3.5}}
\put (4,1) {\line (0,1){3.5}}
\put (6,1) {\line (0,1){3.5}}
\put (8,1) {\line (0,1){3.5}}
\put (11.5,1) {\line (0,1){3.5}}
\put (4,3) {\vector (1,0) {2}}
\put (6,3) {\vector (-1,0) {2}}
\put (7,4.25) {\vector (1,0) {0.5}}
\put (8.5,4.25) {\vector (-1,0) {0.5}}
\put(4.5,3.25){$\frac{T_{\rm \pi}}{n}$}
\put(8.25,4.5){$\tau_{\rm p}$}
\put(11.75,4.0){probe pulse}
\put(8.35,1.75){$\pi$ pulse}
\put(14.25,0.75){$t$}
\put(9.35,1.5){...}
\put(1.75,0.25){0}
\put(11.35,0.25){$T_{\rm \pi}$}
\put(1,5.2){$\Omega$}
\end{picture}
{\caption{Probe pulses and $\pi$ pulse}}
\end{figure}
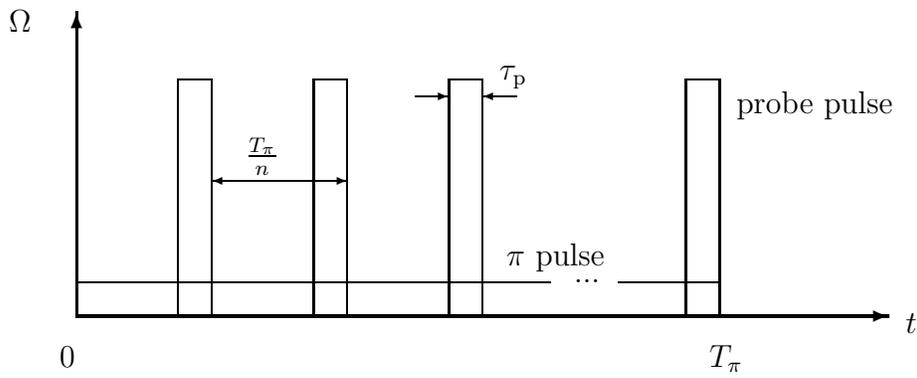

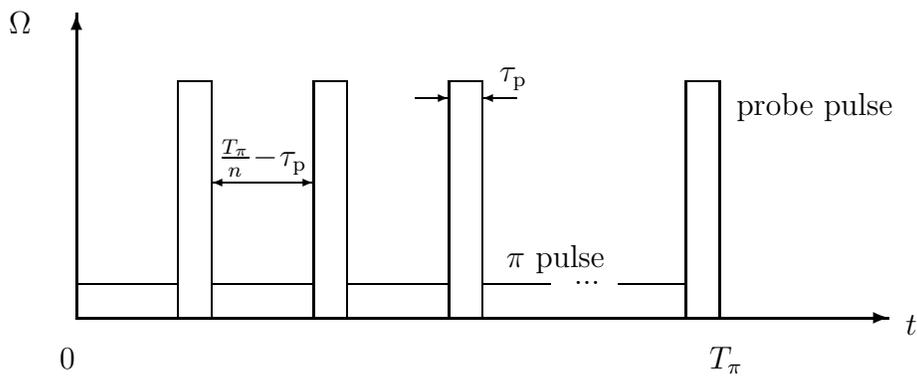
\begin{figure}
\unitlength 0.9cm
\begin{picture}(15.5,6.5)
\thicklines
\put (2,1) {\vector (1,0) {12}}
\put (2,1) {\vector (0,1) {4.5}}

\thinlines
\put (3.5,4.5) {\line (1,0){0.5}}
\put (5.5,4.5) {\line (1,0){0.5}}
\put (7.5,4.5) {\line (1,0){0.5}}
\put (11,4.5) {\line (1,0){0.5}}
\put (2,1.5) {\line (1,0){1.5}}
\put (4,1.5) {\line (1,0){1.5}}
\put (6,1.5) {\line (1,0){1.5}}
\put (8,1.5) {\line (1,0){1.0}}
\put (10,1.5) {\line (1,0){1.0}}
\put (3.5,1) {\line (0,1){3.5}}
\put (5.5,1) {\line (0,1){3.5}}
\put (7.5,1) {\line (0,1){3.5}}
\put (11,1) {\line (0,1){3.5}}
\put (4,1) {\line (0,1){3.5}}
\put (6,1) {\line (0,1){3.5}}
\put (8,1) {\line (0,1){3.5}}
\put (11.5,1) {\line (0,1){3.5}}
\put (4,3) {\vector (1,0) {1.5}}
\put (5.5,3) {\vector (-1,0) {1.5}}
\put (7,4.25) {\vector (1,0) {0.5}}
\put (8.5,4.25) {\vector (-1,0) {0.5}}
\put(4.1,3.25){$\frac{T_{\rm \pi}}{n}\!-\!\tau_{\rm p}$}
\put(8.25,4.5){$\tau_{\rm p}$}
\put(11.75,4.0){probe pulse}
\put(8.35,1.75){$\pi$ pulse}
\put(14.25,0.75){$t$}
\put(9.35,1.5){...}
\put(1.75,0.25){0}
\put(11.35,0.25){$T_{\rm \pi}$}
\put(1,5.2){$\Omega$}
\end{picture}

{\caption{$\pi$ pulse switched off during probe pulse}}
\end{figure}


\begin{thebibliography}{99}
\bibitem{pulse} A $\pi$ pulse, here an rf pulse, transforms the initial
state $|1\rangle$ into $|2 \rangle $ at the end of the pulse, if no
measurements are performed.
\bibitem{obvious} In particular the latter part is not quite obvious
for a superposition. This will be discussed in Section 3.
\bibitem{Gag2} Ref. \cite{Gag} also uses extensive numerical
simulations of quantum trajectories to gain insight into the
evolution of the atomic-level population when the system is
continuously observed. We, on the other hand, use quantum
trajectories in an analytic way.
\bibitem{proof} This is seen as follows. The second condition
means that on average several photons should be
produced during a probe pulse. Estimating  the number of photons by the
steady-state solution of the Bloch equations \cite{Loudon} for
the $1 \leftrightarrow 3$ transition, driven by the probe pulse and
with no population in level 2,
 gives the condition
\begin{eqnarray}
A_3~\frac{\Omega^2_3}{A_3^2 + 2 \Omega^2_3}~\tau_{\rm p}~ >~ 1~,
\nonumber
\end{eqnarray}
where $\Omega_3$ is the Rabi frequency of the probe pulse.
This last condition now implies
\begin{eqnarray}
\tau_{\rm p} > ~\frac{A_3}{\Omega_3^2}~ + ~\frac{2}{A_3} \nonumber
\end{eqnarray}
and hence, by Eq. (\ref{1.10}),
\begin{eqnarray}
T_{\rm \pi} = ~\frac{\pi}{\Omega_2}~ \gg~\frac{A_3}{\Omega^2_3}
{}~+~\frac{2}{A_3}~. \nonumber
\end{eqnarray}
This implies the first part of Eq. (\ref{1.15}). The last equation
can also be written as
\begin{eqnarray}
\frac{\Omega_2}{\Omega_3}~ \left( \frac{A_3}{\Omega_3}~ +
2 ~\frac{\Omega_3}{A_3} \right)~\ll~\pi ~. \nonumber
\end{eqnarray}
Using the inequality  $x + 2/x \ge 2 \sqrt{2} $ for $x > 0$
this gives the condition on
the Rabi frequencies in Eq. (\ref{1.15}).
This condition is also directly obtained by noting that level 1 has
to be pumped at least once to level 3, and hence $\tau_{\rm p} >
{}~\pi/\Omega_3$. Eq. (\ref{1.10}) then implies $\epsilon_{\rm R} \ll
1$. This condition also automatically ensures the obvious
requirement that during a probe pulse the $\pi$ pulse should have
negligible effect.
\bibitem{general}The case with a general $\Omega_3$ will be treated
elsewhere.
\bibitem{ensemble} By an ensemble we always mean a quantum mechanical
ensemble generated by (infinite) repetition  of the preparation of
single systems or by
means of a collection ("gas") of noninteracting systems without cooperative
effects.
\bibitem{jump} The term "quantum jump" is ubiquitous in physics. E.g.,
Cook \cite {B2} defines it as the switching of the energy
eigenvalue under frequent ideal measurements and  uses it to
calculate the density matrix of a two-level system for n ideal
measurements by the projection postulate. It is therefore better to
speak of the "quantum jump approach in quantum optics".
\bibitem{red} Note that this is applied here to photon measurements
only, not to atomic measurements. These reductions can be avoided
\cite{HeWi}  by using the well-known Markov property which is also
employed in the derivation of the Bloch equations as in
 C. Cohen Tannoudji, J. Dupont-Roc, and G. Grynberg,
               {\em Atom-Photon Interactions}, Wiley-Interscience (New
                York 1992).
\bibitem{Norm}  Depending on the parameters  there may be a finite
probability that no photon is emitted at all. Therefore  this probability
density need not be normalized to 1.
\bibitem{end} A small
transient time  after the switch-off of the
probe pulse to allow for the decay of a possible 3-component is
included..
\bibitem{id} For this one may use the identities
$\sum \lambda_i = {\rm tr}\, M = \frac{1}{2} A_3$,
$ \lambda_1 \lambda_2 + \lambda_2 \lambda_3 + \lambda_3 \lambda_1
= \frac{1}{4} (\Omega_2^2 + \Omega_3^2)$,
and $p(\lambda_2) = 0$.
\bibitem{photon}
With probability $\epsilon_{\rm R}^2$ a photon may
be emitted by an atom during this transient time. Such an atom
leaves the subensemble under consideration.
\bibitem{remark} In the experiment of Ref. \cite{Wine} the decay rate of
        the third level is  $1.2 \cdot 10^8 /{\rm s}$. The Rabi frequency
        $\Omega_3$ is given by $1.9 \cdot 10^6 /{\rm s}$, with
        a photon rate of $72$ photons per probe pulse measured in steady
        state.  $\Omega_2$ is given by $12.27 /{\rm s}$, $\tau_{\rm p} =
2.4 \cdot 10^{-3}~{\rm s}$, $T_\pi = .256~{\rm s}$, $\epsilon_{\rm p} =
 4.08 \cdot
10^{-4}$, and $\epsilon_R = 6.46\cdot 10^{-6}.$
\bibitem{rem3} The  same corrections can also be obtained by an
expansion in powers of $\Omega_2$ up to second order as follows. For
an operator $L$ of the form  $L = L_0 + L_1$ one has the identity
$$
\exp{Lt} = \exp{L_0t} + \int_0^t d\tau \exp{L_0(t-\tau)}~L_1
\exp{L\tau}
$$
as checked by differentiation.
This can be iterated and applied to $\rho^0$ and to the complete
density matrix. In both cases the corresponding $L_1$ is proportional
to $\Omega_2$.
\bibitem{rem4} Without stated explicitly, these approximate equations
may also have been used in Ref. \cite{Wine}.
\end{thebibliography}
\end{document}